\documentclass[aps,preprint,tightenlines,showkeys,nofootinbib,superscriptaddress]{revtex4}

\usepackage{graphicx}  %
\usepackage{amsmath}
\usepackage{bm}  %
\usepackage{dsfont}
\usepackage{ulem}

\newcommand{\bea}{\begin{eqnarray}}
\newcommand{\eea}{\end{eqnarray}}
\newcommand{\beq}{\begin{equation}}
\newcommand{\eeq}{\end{equation}}
\newcommand{\bqa}{\begin{eqnarray}}
\newcommand{\eqa}{\end{eqnarray}}

\def\mqo2{{\!\!\!}}

\begin{document}

\title{Charge form factors of two-neutron halo nuclei in halo EFT}
\author{P. Hagen}
\author{H.-W. Hammer}
\affiliation{Helmholtz-Institut f\"ur Strahlen- und Kernphysik and Bethe Center for Theoretical Physics, Universit\"at Bonn, 53115 Bonn, Germany\\}
\author{L. Platter}
\affiliation{Argonne National Laboratory, Physics Division, Argonne, IL 60439,
USA}
\affiliation{Department of Fundamental Physics, Chalmers University of Technology, 
SE-412 96 Gothenburg, Sweden\\}
\date{\today}

\begin{abstract}
We set up a formalism to calculate the charge form factors 
of two-neutron halo nuclei with S-wave neutron-core interactions
in the framework of the halo effective field theory.
The method is applied to some known and suspected halo nuclei.
In particular, we calculate the form factors and charge radii relative 
to the core
to leading order in the halo EFT and compare to experiments where they
are available. Moreover, we investigate the general dependence of the
charge radius on the core mass and the one- and two-neutron
separation energies.
\end{abstract}

\smallskip
\maketitle

\section{Introduction}
\label{sec:intro}
\noindent
The determination of properties of nuclei along the neutron drip line
poses one of the major challenges for modern nuclear experiment and
theory. The associated observables are an important input to studies
of stellar evolution and the formation of elements and provide
insight into fundamental aspects of nuclear
structure. Since these systems are weakly-bound, drip-line nuclei
display novel phenomena associated with newly emerging degrees of 
freedom or, phrased differently, strong correlations. 

Halo nuclei are one example where the transition
to new degrees of freedoms becomes especially apparent. These
nuclei have a tightly bound core with weakly-attached valence
nucleons~\cite{Riisager-94,Zhukov-93,Hansen-95,Jensen-04}.
An up to date overview of the experimental and theoretical state of the
art in the field of halo nuclei can be found in the proceedings of
a recent Nobel Symposium on physics with radioactive beams \cite{NobelT152}.
Usually, halo nuclei are identified by an extremely large matter
radius or a sudden decrease in the one- or two-nucleon separation
energy along an isotope chain. Thus they display a separation of
scales which exhibits itself also in low-energy scattering observables
through a scattering length $a$ that is large compared to the range of
the nucleon-nucleus interaction $R$. Halo nuclei can be
studied with an effective field theory (EFT) that
exploits this separation of scales as a small expansion parameter
$R/a$ and is formulated in the relevant degrees of freedom
\cite{Bertulani-02,BHvK2} (See, e.g., Refs.~\cite{Platter:2009gz,Hammer:2010kp}
for recent reviews.) In
this EFT, the core and the spectator particles are treated as the
fundamental fields in the problem and the overall computational
complexity decreases significantly. In contrast to ab initio
approaches which try to predict nuclear observables from a 
fundamental nucleon-nucleon interaction, halo EFT essentially provides 
relations between different nuclear low-energy observables.
On the one hand, it thus provides a framework that
facilitates a consistent calculation of continuum and bound state
properties when information on the nucleon-nucleus interaction is
known. On the other hand, halo EFT can be used to determine two-body
scattering properties from few-body observables if a sufficient number
of them is known.

In a previous work, Canham and Hammer
\cite{Canham:2008jd,Canham:2009xg} explored the universal properties
and structure of $2n$ halo nuclei to next-to-leading order (NLO) in the
expansion in $R/|a|$ by describing the halo nuclei as an effective
three-body system consisting of a core and two loosely bound valence
neutrons. Their main focus was the possibility of such three-body
systems to display the universal Efimov effect \cite{Efimov-70} and on
the structure of the halo candidates.  In particular, the matter
density form factors and mean square radii were calculated.
Using this framework,
Acharya et al. recently carried out a detailed analysis of the implications of a matter-radius measurement~\cite{Tanaka:2010}
for the binding energy and existence of excited Efimov states in $^{22}$C \cite{Acharya:2013aea}. 
For a selection of previous studies of the possibility of the
Efimov effect in halo nuclei using three-body models,
see Refs.~\cite{Fedorov-94,Amorim-96,Mazumdar-00,Yamashita:2004pv}.
A recent review can be found in \cite{Frederico:2012xh}.

Halo EFT has also been extended to include Coulomb effects \cite{Higa:2008dn}
and electromagnetic currents \cite{Hammer:2011ye,Rupak:2011nk,Rupak:2012cr,Acharya:2013nia} as well as coupled channel effects \cite{Lensky:2011he}
in two-body halo nuclei.
In this paper, we extend these studies to three-body halo nuclei.
We consider the electromagnetic charge form factor and charge radius
of two-neutron halo nuclei interacting through resonant S-wave
interactions. While these quantities have not been measured yet for
current S-wave halo candidates, we anticipate that the
charge radius will be accessible in the near future due to the
constant experimental progress in laser spectroscopy techniques on
radioactive ions that facilitate a measurement of the atomic isotope
shift.

This manuscript is organized out as follows: In Secs.~\ref{sec:lagrangian}, \ref{sec:dimer_renormalization}
and \ref{sec:t_matrix}, we lay out the field theoretical formalism required for the calculation of 
strong interaction observables in two- and three-body S-wave halos. The trimer Greens functions
are introduced in Sec.~\ref{sec:trimer_renormalization} and the calculation of the charge form factor 
is discussed in Sec.~\ref{sec:electric_formfactors}. Our results for the form factors and
radii including error estimates are presented in Sec.~\ref{sec:results}. Finally, we conclude in Sec.~\ref{sec:conclusion}.
Detailed expressions for the various contributions to the form factor are given in the Appendix.

\section{Effective Langrangian}
\label{sec:lagrangian}

\noindent


We set up a non-relativistic effective field theory for a core ($c$) with spin $0$,
mass $m_0$ and electromagnetic charge $\mathcal Z e$, 
interacting with two neutrons ($n$)  with spin $1/2$ and mass $m_1$.  The core is 
described by a scalar field $\psi_0$ and the neutrons are described by a two-component spinor field
$\vec \psi_1 = \left(\begin{smallmatrix} n_\uparrow \\ n_\downarrow \end{smallmatrix}\right)$.


We assume that all two-particle interactions are short-ranged and dominated by S-wave resonances. 
If they are non-resonant, higher-partial wave interactions are suppressed by three powers 
of $R/a$. An EFT formalism for the treatment of resonant interactions in higher partial waves was 
developed in Refs.~\cite{Bertulani-02,BHvK2} and applied to $^6$He in Refs.~\cite{Rotureau:2012yu,Ji:2011}. 
The extension of our form factor formalism to these cases will be left for future work. 
The core-neutron interaction is described by a
spin-$1/2$ dimer field $\vec d_1 = \left(\begin{smallmatrix} d_{1,\uparrow} \\ d_{1,\downarrow} 
\end{smallmatrix}\right)$ and the interaction of the two neutrons is described by a 
spin-$0$ dimer-field $d_0$. The labeling and indices are chosen to simplify the notation
for the three-body equations derived below.
Moreover, we allow for a three-body contact interaction between
the core and the two neutrons which is mediated by a spin-$0$ trimer auxiliary field $t$. Note, that
our choice to introduce auxiliary fields does not imply bound states
in the corresponding channels and merely is a convenient way to introduce interactions.


In addition to the strong interactions between the neutrons and the core, we include
electromagnetic interactions with a vector potential $A_\mu$. The interaction
terms are obtained by minimal coupling which insures gauge invariance:
$i\partial_\mu \mapsto i\partial_\mu - \hat Q A_\mu$, where $\hat Q$ is the charge operator.
In our case only the core has non-vanishing charge ${\mathcal Z}e$, such that 
$\hat Q \psi_0 = \mathcal Z e\, \psi_0$ and  $\hat Q  \vec \psi_1 = 0$ holds. 
For convenience,  we choose Coulomb gauge where $(\vec \nabla \cdot \vec A) = 0$.
Since we restrict our analysis to leading order (LO), non-minimal coupling terms do not
contribute. The effective Lagrangian can then be 
written as the sum of one-, two- and three-body contributions,
$\mathcal L \ = \ \mathcal L^{(1)} \ + \ \mathcal L^{(2)} \ + \ \mathcal L^{(3)}$, where
\begin{equation}
  \begin{split}
    \mathcal L^{(1)} \ =& \ \psi_0^\dagger \left(i\partial_0 + \frac{\vec \nabla^2}{2m_0}\right)\psi_0 \ 
    + \ \vec \psi_1^\dagger \left(i\partial_0 + \frac{\vec \nabla^2}{2m_1}\right) \vec \psi_1 \ 
    - \ \mathcal Z e \ \psi_0^\dagger A_0 \,\psi_0 \\
    &- \frac{1}{2m_0}\psi_0^\dagger \left( i2\hat Q \vec A\cdot \vec \nabla + \hat Q^2 \vec A^2\right)\psi_0\ ,
    \\
    \mathcal L^{(2)} \ =& \ \Delta_1 \, \vec d_1^{\;\dagger} \vec d_1 \ 
    - \ g_1\left[\vec d_1^{\;\dagger} \vec \psi_1 \, \psi_0 + \psi_0^\dagger \, \vec \psi_1^\dagger \vec d_1\right] \ 
    \\
    &+ \ \Delta_0 \, d_0^\dagger  d_0 \ - \ \frac{g_0}{2}\left[d_0^\dagger \, 
        (\vec \psi_1^\text{\;T} P \,\vec \psi_1) + (\vec \psi_1^\text{\;T} P \,\vec \psi_1)^\dagger \, d_0\right] \ ,
    \\
    \mathcal L^{(3)} \ =& \ \Omega \, t^\dagger \, t \ - \ h\left[t^\dagger \, \psi_0\,d_0 + (\psi_0\,d_0)^\dagger t\right] \ .
  \end{split}
  \label{eq:lag_05}
\end{equation}
Because we focus on resonant S-wave interactions, the electromagnetic interaction appears only in $\mathcal L^{(1)}$.


The two-body part $\mathcal L^{(2)}$ includes the bare dimer propagators and the coupling of a dimer to two single particles. 
The bare parameters $\Delta_0$, $g_0$, $\Delta_1$ and $g_1$ depend on the ultraviolet cutoff $\Lambda$. 
At LO the parameters $\Delta_i$ and $g_i$ ($i=0,1$) are not independent. Physical observables only depend on the
the combination $g_i^2/\Delta_i$. The 
spin projection matrix $P$ projects the two neutrons on the spin-singlet. Its components are the corresponding Clebsch-Gordon 
coefficients, leading to
\begin{equation}
  P \ = \ \frac{1}{\sqrt{2}}\left(\begin{smallmatrix}0 & 1 \\ -1 & 0 \end{smallmatrix}\right) \ = \ -P^\dagger \ , 
  \label{eq:lag_20}
\end{equation}
such that $PP^\dagger \ = \ \mathds{1}_2/2$ and $\text{Tr}[PP^\dagger] \ = \ 1$.

Finally, $\mathcal L^{(3)}$ represents the three-body interaction written in terms of a trimer
auxilliary field (see also Ref.~\cite{Bedaque:2002yg}). It includes
the bare trimer propagator and the coupling of the trimer $t$ to the $d_0$-dimer and the 
core field $\psi_0$. Writing the three-body interaction using a trimer 
auxilliary field will be convenient for deriving the form factor expressions below.
The bare parameters $\Omega$ and $h$ depend on the ultraviolet cutoff $\Lambda$.
Again only the combination $h^2/\Omega$ contributes to observables at LO. 


There exists a whole class of equivalent theories in the three particle sector.
Integrating out the auxiliary fields, one can show that different choices of $\mathcal L^{(2)}$ 
and $\mathcal L^{(3)}$ can be transformed into the same theory without dimer and trimer fields up
to four- and higher-body interactions.
To demonstrate this, we eliminate the trimer field $t$ using the classical equation of motion, resulting in
\begin{equation}
  \mathcal L^{(3)} \, \mapsto \, -H_0(\psi_0 d_0)^\dagger(\psi_0 d_0) \ ,
  \label{eq:lag_25}
\end{equation}
where $H_0 \ = \ \frac{h^2}{\Omega}$.
Repeating this step for the dimer fields $\vec d_1$ and $d_0$ then yields
\begin{equation}
  \begin{split}
    \mathcal L^{(2)} \, \mapsto& \, -C^1_0(\psi_0 \vec \psi_1)^\dagger(\psi_0 \vec \psi_1) 
    - C^0_0 (\vec \psi_1^\text{\,T} P \,\vec \psi_1)^\dagger (\vec \psi_1^\text{\,T} P \vec \psi_1) \ , 
    \\
    \mathcal L^{(3)} \, \mapsto& 
     - H'_0\,\left( \psi_0 \, (\vec \psi_1^\text{\,T}P\vec \psi_1) \right)^\dagger \, 
       \left(\psi_0 \, (\vec \psi_1^\text{\,T}P\vec \psi_1)\right)  \ 
       + \  \mathcal L^{(\geq4)} \ , 
  \end{split}
  \label{eq:lag_30}
\end{equation}
where  $C^1_0 \ = \ \frac{g_1^2}{\Delta_1}$, $C^0_0 \ = \ \frac{ g_0^2}{4\Delta_0}$, and
$H'_0 \ = \ \frac{C^0_0 H_0}{\Delta_0}$. The term $\mathcal L^{(\geq4)}$ 
includes interactions of four or more particles. In this work, 
we will only consider processes with at most three particles and therefore neglect $\mathcal L^{(\geq4)}$. 
By this procedure, physical observables will be unchanged
as long as the coupling constants are chosen appropriately.
In particular, the trimer field could also have been introduced in another channel, such as
\begin{equation} 
\mathcal{\tilde{L}}^{(3)} = \Omega \, t^\dagger \, t \ - \ h\left[t^\dagger \, (\vec \psi_1^\text{\,T} P \vec d_1) 
+ (\vec \psi_1^\text{T} P \vec d_1)^\dagger 
t\right]\ ,
\end{equation}
without changing any three-body observables.


In order to write down the derived expressions more compactly, we define the mass parameters:
\begin{equation}
  \begin{split}
    &M_\text{tot} = m_0+2m_1 \ , \qquad M_i=M_\text{tot}-m_i \ , \qquad \mu_i = \frac{m_0m_1^2}{m_iM_i} \ , 
\qquad \tilde \mu_i = \frac{m_i M_i}{M_\text{tot}} \ .
  \end{split}
  \label{eq:lag_40}
\end{equation}
In the following, we use Feynman rules in momentum space to calculate the properties of
the $cn$, $nn$, and $cnn$ systems. In the Feynman diagrams, particles, dimers and trimers 
are denoted by single, double and triple lines, respectively. In addition, propagators are represented by arrows, 
photon couplings by rectangles and all other couplings by ellipses. These symbols are empty if they 
correspond to bare and filled if they correspond to full, interacting quantities.
Since we consider a non-relativistic theory, the one-body properties are not modified by interactions.
We thus start with the two-body problem in the next section.

\section{Two-body problem}
\label{sec:dimer_renormalization}

\noindent


\begin{figure}[ht]
  \centerline{ \includegraphics*[angle=0,clip=true]{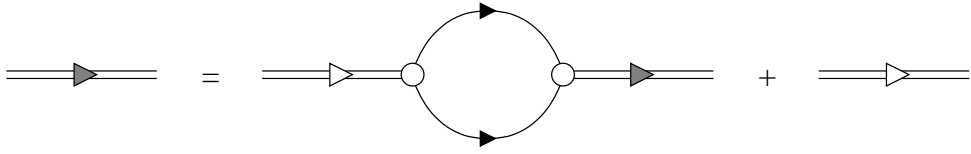} }
  \vspace*{0.0cm}
  \caption{Diagrammatic representation of the integral equation for the full dimer propagator. 
           Particles and dimers are denoted by single and  double lines, respectively. The bare (full) dimer
           propagators are indicated by the empty (filled) arrow.}
  \label{fig:FG_2ps_full_dimer}
\end{figure}


We start by calculating the full dimer propagator $D_i(p_0,\textbf{p})$ with three-momentum $\textbf{p}$ and energy $p_0$ 
for the dimer field with index $i=0,1$, corresponding to the $nn$ and $cn$ channels, respectively.
The integral equation is depicted in Fig.~\ref{fig:FG_2ps_full_dimer} in terms of Feynman diagrams. Using the Feynman rules
derived from the effective Lagrangian (\ref{eq:lag_05}), we find
\begin{equation}
  \begin{split}
    iD_i(p_0, \textbf{p} ) \ =& \ \frac{i}{\Delta_i} \ \left(-i\Sigma_i(p_0, \textbf{p} ) \right) \  iD_i(p_0, \textbf{p} ) \ 
    + \ \frac{i}{\Delta_i} \ = \ i\left[\Delta_i - \Sigma_i(p_0, \textbf{p} ) \right]^{-1} \ , 
    \\
    -i\Sigma_i(p_0, \textbf{p} ) \ =& \ i\frac{s_i\,g_i^2\mu_i}{\pi^2} \left[ \Lambda  - \frac{\pi}{2} \, \sqrt{2\mu_i
  \left(\frac{\textbf{p}^2}{2M_i} - p_0 - i\varepsilon\right)} \right]\ ,
  \end{split}
  \label{eq:2ps_10}
\end{equation}
where $s_i=\delta_{i0}/2 + \delta_{i1}$ is a symmetry factor and $-i\Sigma_i(p_0, \textbf{p} )$ is the 
dimer self energy. It is linearly divergent and has to be regularized. 
For this purpose, we choose a  momentum cutoff $\Lambda$ which is the same in both channels.
Contributions to the self-energy suppressed by powers of $1/\Lambda$ have been omitted
in Eq.~(\ref{eq:2ps_10}). They are small for large $\Lambda$ and can be absorbed in the 
renormalized coupling constants.


Matching the two-body T-matrix obtained from the propagator (\ref{eq:2ps_10})
for $\textbf{p}=0$ and $p_0=k^2/(2\mu_i)$
to the effective range expansion of the S-wave scattering amplitude
\begin{equation}
f_i(k) \ = \ \left[ -\frac{1}{a_i} +\mathcal O (k^2) -ik\right]^{-1}
  \label{eq:2ps_20}
\end{equation}
with scattering length $a_i$, we eliminate the dependence on the cutoff $\Lambda$. 
This leads to the renormalization condition for the two-particle couplings
\begin{equation}
 \frac{1}{a_i} \ = \ \frac{2\pi\Delta_i}{s_i \,g_i^2\mu_i} + \frac{2}{\pi} \Lambda
  \label{eq:2ps_30}
\end{equation}
and the renormalized, full dimer propagator
\begin{equation}
  D_i(p_0, \textbf{p} ) \ = \ \frac{2\pi}{s_i \,g_i^2\mu_i} \ \left[ \frac{1}{a_i} 
- \sqrt{ 2\mu_i\left( \frac{\textbf{p}^2}{2M_i} - p_0 - i\varepsilon\right) } \right]^{-1} \ .
  \label{eq:2ps_40}
\end{equation}
For positive scattering length $a_i$, this propagator has one pole on the first Riemann sheet of 
the complex square root with the positive residue
\begin{equation}
  Z_i \ = \ \frac{2\pi}{s_i \, g_i^2\mu_i^2} \, \frac{1}{a_i} \ .
  \label{eq:2ps_50}
\end{equation}
This pole corresponds to a two-body bound state with binding energy $B_i=1/(2\mu_i a_i^2)$. 
For negative $a_i$ there is a pole with negative residue on the unphysical, second Riemann sheet.

The leading correction to the propagator (\ref{eq:2ps_40}) is due to the effective range.
It could be included by making the dimer fields dynamical as discussed, e.g., 
in Refs.~\cite{Kaplan:1996nv,Bedaque:1997qi,Hammer:2001gh,Ji:2011qg}.
Here, we stay at leading order in the EFT expansion and neglect effective range corrections.

\section{Three-body problem}
\label{sec:t_matrix}

\noindent


\begin{figure}[ht]
  \centerline{ \includegraphics*[angle=0,clip=true]{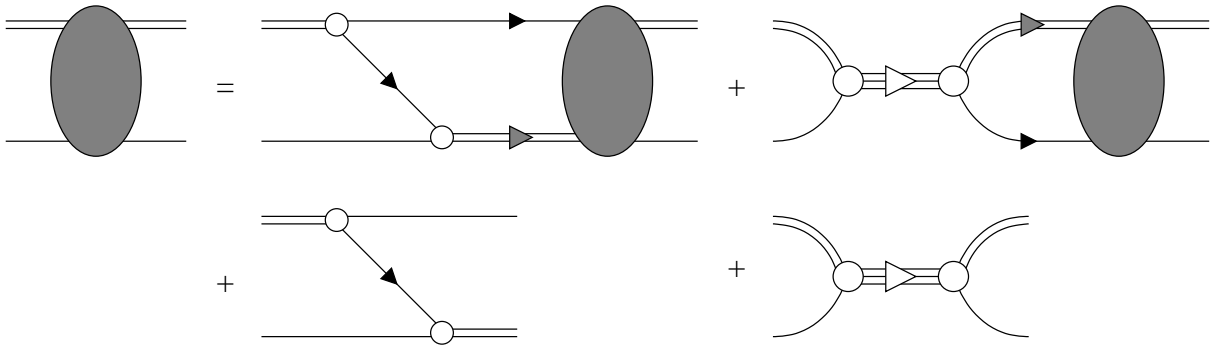} }
  \vspace*{0.0cm}
  \caption{Diagrammatic representation of the integral equation for the particle-dimer
           $T$-matrix. Particles, dimers, and trimers are denoted by single, double, and triple lines, respectively.
           The bare (full) propagators are indicated by the empty (filled) arrows.}
  \label{fig:FG_3ps_full_T-matrix}
\end{figure}


We proceed by calculating the $T$-matrix element $T_{ij}$ for the scattering process of a dimer and a particle.
The dimer and particle in the incoming (outgoing) channel are both labeled by the index $i$ ($j$), respectively.
We focus on the dimer-particle $T$-matrix as a central quantity, since all other three-body observables can
be obtained from it.
The integral equation for the $T$-matrix is depicted in Fig.~\ref{fig:FG_3ps_full_T-matrix} in terms of
Feynman diagrams. The matrix structure of the equation is implicit in  Fig.~\ref{fig:FG_3ps_full_T-matrix}. 
We note that the universal properties and structure of $2n$ halo nuclei were previously explored
in an effective quantum mechanics framework  \cite{Canham:2008jd,Canham:2009xg}. In this work, the 
cluster wave functions were obtained by solving the Faddeev equations for an effective potential
reflecting the expansion in $R/|a|$. Here, we obtain all observables directly from the $T$-matrix.

The derivation of the corresponding expressions can be performed in several steps. First, we project in- and outgoing
states onto the spin-singlet using the projection operator from Eq.~\eqref{eq:lag_20}. 
External dimer fields are then renormalized with the wave function renormalization factors from 
Eq.~\eqref{eq:2ps_50} via 
\begin{equation}
  T_{ij} \ \mapsto \ |Z_i|^\frac12 \, T_{ij} \, |Z_j|^\frac12 \quad .
  \label{eq:3ps_10}
\end{equation}
The absolute values in Eq.~(\ref{eq:3ps_10}) are only required for $i,j=0$ because $Z_0<0$.
This channel corresponds to a neutron-neutron dimer which is unbound and requires no wave function renormalization
factor. In this case, Eq.~\eqref{eq:3ps_10} simply provides a convenient redefinition
of the amplitude but has no physical significance.


We work in the center-of-mass frame, in which the on-shell $T$-matrix only depends on the total energy $E$ and the relative 
momenta in the ingoing and outgoing channels $\textbf{p}$ and $\textbf{k}$, respectively. 
If the dimer and the particle have ingoing (outgoing) three-momenta $\textbf{p}_1$ and $\textbf{p}_2$ ($ \textbf{k}_1$ 
and $ \textbf{k}_2$) in a general frame, the relative momenta are
\begin{equation}
  \textbf{p} \ = \ \frac{m_i}{M_\text{tot}}\textbf{p}_1 - \frac{M_i}{M_\text{tot}} \textbf{p}_2 \ , 
\qquad \textbf{k} \ = \ \frac{m_j}{M_\text{tot}}\textbf{k}_1 - \frac{M_j}{M_\text{tot}} \textbf{k}_2 \ .
  \label{eq:3ps_20}
\end{equation}

The $T$-matrix can be decomposed into partial wave contributions $T^{[\ell m, \ell'm']}$.
The Wigner-Eckart theorem then implies that $T$ is diagonal in  $\ell$ and $m$ and can be written as
$T^{[\ell m, \ell'm']}=\delta_{\ell\ell'}\delta_{mm'}T^{[\ell]}$. 
The resulting $2\times2$-matrix integral equation for angular momentum $\ell$ is a generalization of the 
Skorniakov-Ter-Martyrosian  (STM) equation~\cite{STM:1957} and reads
\begin{equation}
  T^{[\ell]}(E,p,k) \ = \ \int_0^\Lambda \text{d}q \ R^{[\ell]}(E,p,q) \ \bar{D}(E, q) \ T^{[\ell]}(E,q,k) \ + \ R^{[\ell]}(E,p,k) \ ,
  \label{eq:3ps_30}
\end{equation}
where $\Lambda$ is an ultraviolet cutoff on the loop-momentum in the three-particle sector. The components of the interaction 
matrix $R^{[\ell]}$ are given through
\begin{equation}
  \begin{split}
    R^{[\ell]}_{ij}(E,p,k) \ =& \ \frac{ (1-\delta_{i0}\delta_{j0}) }{ \sqrt{ s_i s_j } } \, (-1)^\ell \ 
\frac{2\pi}{\sqrt{|a_i| |a_j|}} \ \frac{ M_\text{tot}-m_i-m_j}{\mu_i\mu_j} \, \frac{1}{pk} \ Q_\ell\big(c_{ij}(E,p,k)\big)
    \\
    & \ - \delta_{i0} \delta_{j0} \, \delta_{\ell0} \ H \ ,
    \\
     c_{ij}(E,p,k) \ =& \ \frac{M_\text{tot}-m_i-m_j}{pk} \left(\frac{p^2}{2\mu_j} + \frac{k^2}{2\mu_{i}} -E -i\varepsilon\right) \ ,
  \end{split}
    \label{eq:3ps_40}
\end{equation}
where $Q_\ell$ are the analytically continued
Legendre functions of the second kind. In our numerical calculations, we 
will only need
\begin{equation}
Q_0(c)\ =\ \begin{cases}
\text{arctanh}(1/c) &: |c| > 1\ ,\\
\text{arctanh}(c) +i\frac{\pi}{2} &: |c| < 1\ .
\end{cases}
\end{equation}
The form of $Q_0$ is determined by taking the limit $\varepsilon\to 0^+$
in the integral equation.
Moreover, $H=|Z_0|h^2/\Omega$ is the redefined three-body coupling, which 
depends on the cutoff $\Lambda$. It only contributes for angular momentum $\ell=0$. The dimer matrix is diagonal in the
channel indices: 
$\bar{D}= \text{diag}(\bar{D}_0,\bar{D}_1)$ with
\begin{equation}
  \bar{D}_i(E,q) \ = \ \frac{\mu_i |a_i|}{2\pi^2} \ \frac{q^2}{-\frac{1}{a_i} + \sqrt{b_i(E,q)}  } \ , 
\qquad b_i(E,q) \ = \ 2\mu_i\left(\frac{q^2}{2\tilde\mu_i}-E-i\varepsilon\right) \ .
  \label{eq:3ps_50}
\end{equation}
Note that an overall factor $-q^2/(2\pi^2|Z_i|)$ from the measure of the integration over the loop momentum
$q$ and the dimer wave function renormalization
has been absorbed in $\bar{D}_i$ for notational convenience.


Assuming the existence of an S-wave three-body bound state at energy $E=-B$, 
the transition amplitude can be decomposed as
\begin{equation}
  T^{[\ell]}(E, p, k) \ = \ - \delta_{\ell0}\frac{ \vec {\mathcal B}(p) \cdot {\vec {\mathcal B}}^\dagger(k) }{E+B+i\varepsilon}  
\quad + \quad \mbox{regular terms}\ .
  \label{eq:3ps_60}
\end{equation}
The residue of the bound state pole
factors into wave functions $\vec {\mathcal B}(p)$ depending only on one single momentum, 
and the remaining part is a regular function in the energy. Inserting Eq.~\eqref{eq:3ps_60} into Eq.~\eqref{eq:3ps_30}  
yields the bound state equation
\begin{equation}
  \vec {\mathcal B}(p) \ = \ \int_0^\Lambda \text{d}q \ R^{[0]}(E,p,q) \ \bar{D}(E, q) \ \vec {\mathcal B}(q) \ .
  \label{eq:3ps_70}
\end{equation}
This generalized eigenvalue problem has an Efimov-like spectrum of three-body bound state energies. For a given cutoff $\Lambda$, we 
then fix the unknown three-body parameter $H$ such that Eq.~\eqref{eq:3ps_70} has a solution at 
the desired value $E=-B$. 
In this way, the three-body coupling is renormalized and other three-body observables can be predicted.
In particular, Eq.~\eqref{eq:3ps_30} can be solved numerically in order to determine the $T$-matrix for three-body scattering
observables.


In the following, we will consider only three-body observables in the S-wave ($\ell=0$) channel 
and drop the index ``$[0]$'' on the quantities $R^{[0]}$ and $T^{[0]}$ for notational simplicity.
From the $T$-matrix, we can derive the scattering amplitude and scattering length for dimer-particle scattering. 
Since the two-neutron system is not bound, only the element $T_{11}$ 
of the $2\times 2$ $T$-matrix in Eq.~(\ref{eq:3ps_30}) describes a physical scattering 
process, namely the scattering of a neutron from a $cn$ bound state:
\begin{equation}
   T_{11}\left(\frac{p^2}{2\tilde\mu_1} - \frac{1}{2\mu_1 a_1^2},\;p,\;p\right)= \ \frac{2\pi}{\tilde \mu_1}\,
    \frac{1}{p\cot\delta_{cn-n}(p)-ip}
  \label{eq:3ps_75}
\end{equation}
where the reduced masses $\mu_1$ and $\tilde \mu_1$ are defined in Eq.~(\ref{eq:lag_40}).

\section{Trimer Greens Functions}
\label{sec:trimer_renormalization}

\noindent


\begin{figure}[ht]
  \centerline{ \includegraphics*[angle=0,clip=true]{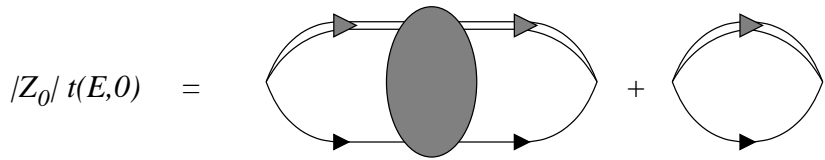} }
  \vspace*{0.0cm}
  \caption{Feynman diagrams contributing to the full trimer propagator $t$, defined with an implicit factor $|Z_0|$.
   Particles, dimers, and trimers are denoted by single, double, and triple 
   lines,  respectively. Otherwise the notation is as in    
   Fig.~\ref{fig:FG_3ps_full_T-matrix}.}
  \label{fig:FG_2ftc_full_trimer}
\end{figure}


In order to calculate the charge form factor of the three-body bound 
states corresponding to two-neutron halo nuclei, we need the trimer 
wave function renormalization $Z_{\text{tr}}$ which is given by the residue 
of the bound state pole in the full trimer propagator including interactions. 
The Feynman diagrams contributing to the propagator
are shown in Fig.~\ref{fig:FG_2ftc_full_trimer}
where the meeting point corresponds to a source for the 
$d_0\psi_0$ state in accordance with our choice for the three-body
interaction in Eq.~(\ref{eq:lag_05}) to act only in the $i=0$ channel.
Thus only $T_{00}$ contributes to $t$.
For a trimer at rest, it can be written as
\begin{equation}
  t(E, 0) \ = \ 
  \int_0^\Lambda \text{d}q \, 
  \int_0^\Lambda \text{d}q'  \, \bar{D}_0(E, q) \, T_{00}( E, q , q' ) \, 
  \bar{D}_0(E, q') 
  \quad + \quad\mbox{regular terms}\ ,
  \label{eq:ftc_20}
\end{equation}
where the $\text{d}q_0$ integrals have already been carried out.
Since in Fig.~\ref{fig:FG_3ps_full_T-matrix}, we defined $t$ with an implicit factor of $|Z_0|$, only renormalized quantities appear in Eq.~(\ref{eq:ftc_20}). Thereby the dimer matrix component $\bar D_0$ from Eq.(~\ref{eq:3ps_50}) comes from the fact that in both loops integrals the single particle propagator sets the loop four-momentum on-shell.
The trimer wave function renormalization  
can be extracted from the relation
\begin{equation}
     Z_\text{tr} \ = \ -\lim_{E\rightarrow-B}(E+B) \, t(E,0) \ .
  \label{eq:ftc_30}
\end{equation}

The trimer self energy is given by all trimer-irreducible contributions to the 
propagator and can be defined as
\begin{equation}
  \Sigma(E) \ = \ t(E,0)\big|_{H=0}\ ,
  \label{eq:ftc_33}
\end{equation}
where the three-body force is set to zero in the evaluation of $T_{00}$.
Using $\Sigma(E)$, the trimer propagator can also be written as
\begin{equation}
  t(E, 0) \ = \ \frac{\Sigma(E)}{1 - H \, \Sigma(E)} \ .
  \label{eq:ftc_35}
\end{equation}
Requiring that $t(E,0)$ has a bound state pole at $E=-B$, Eq.~\eqref{eq:ftc_35} directly leads to the 
relation 
\begin{equation}
\Sigma(-B)=1/H\ .
\label{eq:SigmaB}
\end{equation}

\begin{figure}[ht]
  \centerline{ \includegraphics*[angle=0,clip=true]{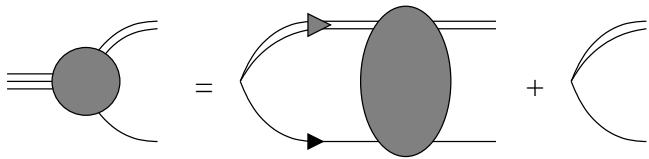} }
  \vspace*{0.0cm}
  \caption{Diagrammatic representation of the irreducible 
   trimer-dimer-particle three-point function $\mathcal G^\text{irr}$.
   The three-body force $H$ is set to zero in the evaluation of the 
   particle-dimer
   $T$-matrix. Notation is as in Fig.~\ref{fig:FG_2ftc_full_trimer}.}
  \label{fig:FG_2ftc_full_G}
\end{figure}
In the explicit expressions for the trimer form factor (cf. Appendix \ref{sec:appedix}), 
we need the irreducible trimer-dimer-particle three-point function defined by
\begin{equation}
    \mathcal G^\text{irr}_i(E,p) \ := 
    \int_0^\Lambda \text{d}q \, \bar{D}_0(E, q) \ T_{0i}(E, q, p)\Big|_{H=0} \ 
     + \ \delta_{0i} \ ,
  \label{eq:ftc_40}
\end{equation}
where again trimer-reducible contributions are removed by setting $H=0$ in the integral equation for $T_{0i}$. 
The corresponding Feynman diagrams are depicted in Fig.~\ref{fig:FG_2ftc_full_G}. Inserting Eq.~\eqref{eq:3ps_30} into Eq.~\eqref{eq:ftc_40} 
and writing $\vec {\mathcal G}^\text{irr} = 
\left(\begin{smallmatrix} \mathcal G^\text{irr}_0 \\ \mathcal G^\text{irr}_1 \end{smallmatrix}\right)$, yields a matrix integral 
equation for the irreducible trimer-dimer-particle three-point function
\begin{equation}
  \vec {\mathcal G}^\text{irr}(E,p) \ = \ \int_0^\Lambda \text{d}q \, R(E, p, q)\Big|_{H=0} \, \bar{D}(E, q) \, 
\vec {\mathcal G}^\text{irr}(E, q) \  + \ \vec {e_0} \ ,
  \label{eq:ftc_45}
\end{equation}
where $(\vec {e_0})_i=\delta_{0i}$. $\vec {\mathcal G}^\text{irr}$ depends on the cutoff $\Lambda$, but the combination 
$\sqrt{Z_{\text{tr}}} \, H \, \vec {\mathcal G}^\text{irr}$ is independent of $\Lambda$ up to an overall sign. This is exactly the
combination that enters into the form factor calculation.

\section{Charge formfactors}
\label{sec:electric_formfactors}


We are now in a position to calculate the charge formfactor $\mathcal F_\text{E}$ of a $cnn$ halo system with resonant S-wave interactions. 
The form factor can be extracted from the matrix element of the electromagnetic current $j_\mu$ between trimer states.
We will denote the in- and outgoing three-momenta of the trimer by $\textbf{P}$ and $\textbf{K}$, respectively. It is convenient 
to extract the charge form factor from the matrix element of the zeroth component of the electromagnetic current 
\begin{equation}
  \left< \, t(K_0, \textbf{K}) \, | \, j_0\, | \, t(P_0, \textbf{P}) \, \right> \ = \ (-ie\mathcal Z) \ 
\mathcal F_\text{E}(\textbf{Q}^2)
  \label{eq:ff_05}
\end{equation}
in the Breit frame, where no energy is transferred by the photon, i.e. $P_0=K_0$ and $\textbf{P}^2= \textbf{K}^2$. 
The charge formfactor in Eq.~\eqref{eq:ff_05}, then depends only on the three-momentum transfer 
$\textbf{Q}^2=(\textbf{K}-\textbf{P})^2$.

The LSZ reduction formula implies that the current matrix element in Eq.~\eqref{eq:ff_05} can be expressed 
as \cite{Kaplan:1998sz}
\begin{equation}
  \begin{split}
    \left< \, t(K_0, \textbf{K}) \, | \, j_0\, | \, t(P_0, \textbf{P}) \, \right> \ =& \ \sqrt{Z_{\text{tr}}} \,  
\Sigma(-B)^{-1} \ i\Gamma_0(\textbf{Q}) \ \Sigma(-B)^{-1} \sqrt{Z_{\text{tr}}}
    \\
    =& \ Z_{\text{tr}} \, H^2 \ i\Gamma_0(\textbf{Q})\ ,
  \end{split}
  \label{eq:ff_15}
\end{equation}
where $i\Gamma_0(\textbf{Q})$ is the sum of all irreducible Feynman diagrams with external trimer lines and 
a photon coupled to the core.

\begin{figure}[ht]
  \centerline{ \includegraphics*[angle=0,clip=true]{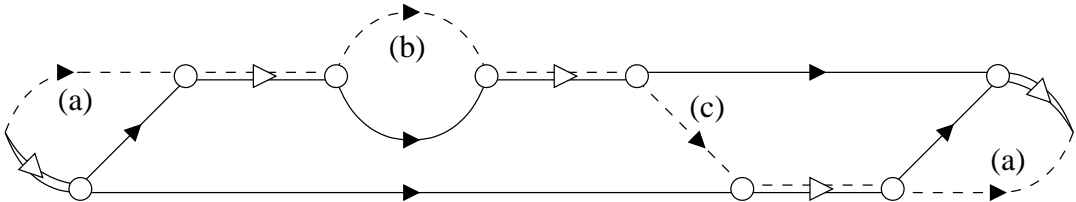} }
  \vspace*{0.0cm}
  \caption{Exemplary irreducible graph contributing to the trimer form factor. 
  The dashed line represents the core field which either 
  (a) propagates parallel to the $d_0$-dimer, (b) appears within a $d_1$-dimer loop or (c) is exchanged between $d_1$-dimers.
   Otherwise the notation is as in Fig.~\ref{fig:FG_2ftc_full_G}.}
  \label{fig:FG_ff_example}
\end{figure}
In order to motivate the different contributions to $i\Gamma_0(\textbf{Q})$, we consider 
the typical irreducible graph shown in Fig.~\ref{fig:FG_ff_example}.
The photon can only couple to the core field $c$ indicated by dashed lines, but
for the moment we suppress the photon-core coupling.\footnote{Of course the $d_1$-dimer also carries charge, but the 
photon coupling to $d_1$ appears only at next-to-leading order where the dimers are dynamical.} 
Within such a diagram, the core (a) propagates 
either parallel to the $d_0$-dimer, (b) appears within a $d_1$-dimer loop, or (c) is exchanged between between 
two $d_1$-dimers. In fact, these are the only 3 possibilities for a single core propagator to appear in an arbitrary 
irreducible trimer graph. Thus, including the photon-core coupling and summing over all such diagrams the form factor 
derived from the transition amplitude through Eq.~\eqref{eq:ff_05} and Eq.~\eqref{eq:ff_15}, can be written as the 
sum of three contributions
\begin{equation}
  \mathcal F_\text{E} \ = \ \mathcal F_\text{E}^{(a)} \ + \ \mathcal F_\text{E}^{(b)} \ + \ \mathcal F_\text{E}^{(c)} \ ,
  \label{eq:ff_30}
\end{equation}
corresponding to the cases (a), (b), and (c). In all three contributions the irreducible trimer-dimer-particle three-point 
function $\vec {\mathcal G}^\text{irr}$ from Eq.~\eqref{eq:ftc_40} appears naturally. 

\begin{figure}[th]
  \centerline{ \includegraphics*[width=\textwidth, angle=0,clip=true]{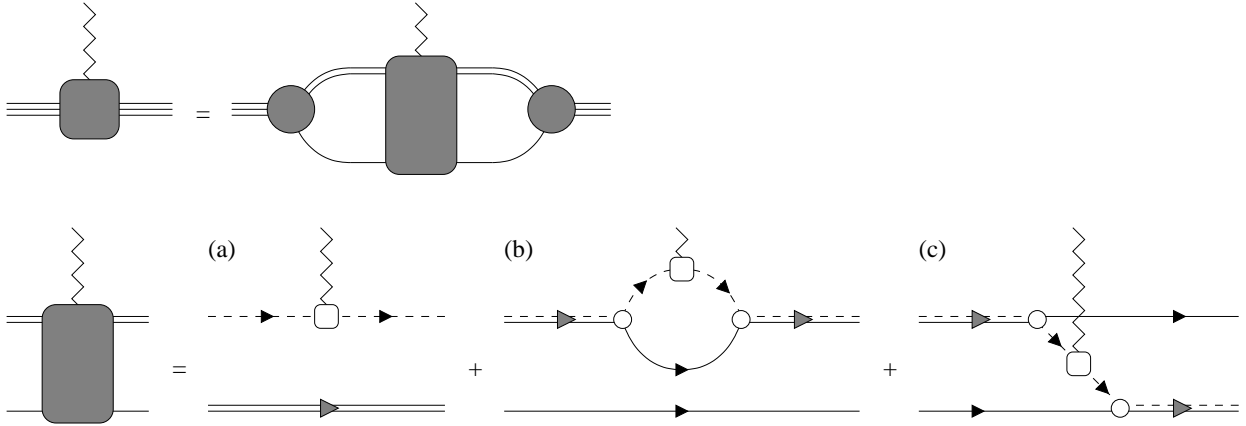} }
  \vspace*{0.0cm}
  \caption{Diagrammatical representation of the form factor matrix element \eqref{eq:ff_05}.
      The contributions fall into three different classes (a), (b), and (c). 
      Notation is as in Fig.~\ref{fig:FG_2ftc_full_G}.}
  \label{fig:FG_ff_contributions}
\end{figure}
In Fig.~\ref{fig:FG_ff_contributions}, 
the decomposition of the form factor matrix element~\eqref{eq:ff_05} into the three classes of diagrams 
is illustrated pictorially. The filled circle represents $\vec {\mathcal G}^\text{irr}$ from Eq.~\eqref{eq:ftc_40}.
Performing shifts in the loop momenta of order $Q=|\textbf{Q}|\ll\Lambda$ one can express $\mathcal F_\text{E}^{(a)}$, 
$\mathcal F_\text{E}^{(b)}$, and $\mathcal F_\text{E}^{(c)}$ through integrals symmetric in the momenta. 
The various contributions are derived in Appendix~\ref{sec:appedix} and explicit expressions are given in 
Eqs.~\eqref{eq:app_09},~\eqref{eq:app_19} and~\eqref{eq:app_29},  respectively.
We note that analog calculations for the charge form factor of the triton (corresponding
to a spin-1/2 core) have previously been carried out
in a wave function based formalism \cite{Platter:2005sj,Sadeghi:2012}.

At $Q^2=0$, the charge formfactor is normalized to one. This
normalization is automatically reproduced in our formalism.
For small momentum transfer, the form factor can be expanded in powers of $Q^2$ 
as:
\begin{equation}
  \mathcal F_\text{E}(Q^2) \ = \ 1 - \frac{\left<r^2_\text{E}\right>}{6} Q^2 \ + \ \mathcal O(Q^4) \ ,
  \label{eq:ff_10}
\end{equation}
where $\left<r^2_\text{E}\right>$ is the charge radius.

In practice, we calculate the form factors for finite momentum transfer and extract the
the charge radius $\left<r^2_\text{E}\right>$ by numerically taking the limit
\begin{equation}
  \left<r^2_\text{E}\right> \ = -6\lim_{Q^2\to0^+}\frac{d
 \mathcal F_\text{E}}{dQ^2}\ .
  \label{eq:ff_50}
\end{equation}
Here, we have to keep in mind, that in our effective theory the core and the neutrons are treated as
pointlike. Their size enters only in counter terms that appear at higher orders. 
In typical halo nuclei, however, the charge radius of the core can not be neglected.
In this work, we thus interpret the calculated radius as the charge radius of the
$cnn$ halo nucleus relative to the charge radius of the core. The small negative charge radius
of the neutron $\left<r^2_\text{E}\right>_n = -0.115(4)$ fm${}^2$~\cite{Kopecky:1997rw}
is neglected. In order to get the full charge radius of the $cnn$ halo nucleus $\left<r^2_\text{E}\right>_{cnn}$,
we therefore have to quadratically add our result to the charge radius of the core $\left<r^2_\text{E}\right>_c$:
\begin{equation}
    \left<r^2_\text{E}\right>_{cnn} \ = \ \left<r^2_\text{E}\right>_c  \ + \ \left<r^2_\text{E}\right>\ ,
 \label{eq:ff_51}
\end{equation}
or simply quote the difference $\delta\left<r^2_\text{E}\right> \ := \ \left<r^2_\text{E}\right>_{cnn} 
- \left<r^2_\text{E}\right>_c $.
This prescription follows directly if the total charge distribution is a convolution of the charge distributions
of the halo and the core.
The differences of nuclear charge radii $\delta\left<r^2_\text{E}\right>$ were measured for a whole range
of isotopes,  see e.g.~\cite{Sanchez:2006, Noertershaeuser:2009, Krieger:2012, Geithner:2008, Yordanov:2012}. 
In the next section, we will compare these results with our theory where it is applicable.

\section{Results}
\label{sec:results}

\noindent

We now apply our effective field theory to concrete physical systems. Our theory applies directly 
to two-neutron halo nuclei with $J^\pi=0^+$ and with a $J^\pi=0^+$ core. Assuming that the spin of the core 
is inert due to the large mass of the core compared to the neutrons, we can also consider more general systems with 
quantum numbers $J^\pi$, $(J\pm1/2)^\pi$, and $J^\pi$ of the $c$-, $cn$-, and $cnn$-systems, respectively. 
From now on, we write $m_n$ for the neutron mass and $m_c$ for the core mass for convenience.

At this point, a discussion of the different types of errors in 
our calculation is in order. There are three types of errors: 
$(i)$ the numerical errors in our calculation which are negligible. 
$(ii)$ errors in the input used to fix the effective
theory parameters. These errors can be propagated to our final results. 
For the case of $^{22}$C, e.g., these errors dominate. 
$(iii)$ errors from higher orders in the
EFT expansion. These errors come from operators that contribute at the next 
order and are difficult to obtain.
Short of an explicit higher order calculation, one must use dimensional 
analysis and naturalness to estimate their size. In most nuclei 
considered in the present paper the errors of type $(iii)$ dominate. 
We note that errors of type $(iii)$ can never be provided in model 
calculations since no expansion scheme exists. In this sense, model 
calculations are uncontrolled.

In order to have a bound or virtual $cn$-dimer and a bound $cnn$-trimer their one- and two-neutron separation energies 
$B_{cn} \equiv B_1$ and $B_{cnn} \equiv B$ have to obey $ B_{cnn} > \text{max}(B_{cn},0)$. Moreover, we denote the first
excitation energy of the core by $E^*_c$ and the one-neutron separation energy by $B_{c-n}$. 
As discussed in the introduction, the expansion parameter of
our theory $R_\text{core}/R_\text{halo}$ is roughly $R/a$. In order to obtain better estimates,
we compare the typical energy scales $E_\text{halo}$ and  $E_\text{core}$ of the neutron 
halo and the core, respectively. To estimate $E_\text{halo}$, 
we choose the one- or two-neutron separation energy $B_{cn}$ or $B_{cnn}$. 
The energy scale of the core is estimated by its excitation energy $E^*_c$ or its one-neutron separation energy $B_{c-n}$. The 
square root of the energy ratio $R_\text{core}/R_\text{halo} \approx \sqrt{|E_\text{halo}/E_\text{core}|}$ then yields an
estimate for the expansion parameter of the effective theory. 
In particular, if $R_\text{halo}$ is estimated from 
$E^*_c$ or $B_{c-n}$ this ratio quantifies the quality of 
the structureless core approximation.
For our error estimates, we take the largest value for $R_\text{core}/R_\text{halo}$ that can be obtained this way. 


We fix the values of all quantum numbers $J^\pi$, masses, and energies by taking data from the  National Nuclear Data Center
(NNDC)~\cite{NNDC:2011} unless noted otherwise.
The $cn$-scattering length is determined from the relation $a_1\equiv a_{cn}=\text{sign}(B_1)/\sqrt{2\mu_1|B_1|}$.
Thus $cn$-states with negative $B_1$ are treated as virtual two-body states with negative scattering length $a_1$.
This approximation corresponds to neglecting the imaginary part of the binding momentum for resonances.
For the $nn$-scattering length, we take the value $a_0\equiv a_{nn}=-18.7(6)$~fm from Gonzales Trotter 
et al.~\cite{GonzalezTrotter:1999}. 


The lightest isotopes for which there is either experimental evidence for their $2n$-halo nature or 
which are good candidates for such a system, are ${}^{6}$He, ${}^{11}$Li, ${}^{14}$Be, ${}^{17}$B and ${}^{22}$C. 
Since the $J^\pi$-quantum numbers of ${}^{6}$He and ${}^{17}$B indicate that P-wave contributions must be dominant,
we apply our effective theory to ${}^{11}$Li, ${}^{14}$Be and ${}^{22}$C. 


In Tab.~\ref{tab:r2_a}, we summarize the effective theory parameters and our predictions for the charge radii 
relative to the core
in ${}^{11}$Li, ${}^{14}$Be and ${}^{22}$C. Uncertainties in the energies are only 
quoted if they are larger than~$1\%$. The small uncertainties in the nuclear masses can be neglected. For ${}^{11}$Li 
and ${}^{14}$Be, the expansion parameter $R_\text{core}/R_\text{halo}$ is typically not much smaller than~$1$. As a 
consequence, the main uncertainty in our calculation for these systems is from the next-to-leading order corrections 
in the effective theory. In the ${}^{22}$C halo nucleus, the main uncertainty is from the poor knowledge of the binding 
energies.  Below, we discuss our analysis for each halo nucleus in detail.


\begin{table}[ht]
  \begin{tabular}{|c|c|c||c|c||c|}
    \hline
    $c$ & $J^\pi_{c}$ & $m_c \quad $ [MeV] & $E_c^*$ [MeV] & $B_{c-n}$ [MeV]& 
    $\delta\left<r^2_\text{E}\right>$ [fm$^2$]
    \\
    $cn$ & $J^\pi_{cn}$ & $B_{cn} \ \, $ [MeV] & $\frac{B_{cn}}{E_c^*}$ & $\frac{B_{cn}}{B_{c-n}}$ 
    & $\delta\left<r^2_\text{E}\right>_\text{exp}$ [fm$^2$]
    \\
    $cnn$ & $J^\pi_{cnn}$ & $B_{cnn}$ [MeV] & $\frac{B_{cnn}}{E_c^*}$ & $\frac{B_{cnn}}{B_{c-n}}$ 
    &
    \\
    \hline
    \hline
    ${}^{9}$Li & $\frac32^-$  & $8406$ & $2.69$ &  $4.06$ & $1.7(6)$
    \\
    ${}^{10}$Li & $(2^-, 1^-)$ & $-0.026(13)$ & $-0.10^2$ & $-0.08^2$ & $1.171(120)$~\cite{Sanchez:2006}
    \\
    ${}^{11}$Li & $\frac32^-$ & $0.37$ & $0.37^2$ & $0.30^2$ &
    \\
    \hline
    ${}^{12}$Be & $0^+$ & $11201$ & $2.10$ & $3.17$ & $0.4(3)$
    \\
    ${}^{13}$Be & $(\frac12^-)$& $-0.51(1)$ & $-0.49^2$ & $-0.40^2$ & $--$
    \\
    ${}^{14}$Be & $0^+$ & $1.27(13)$ & $0.78^2$ & $0.63^2$ &
    \\
    \hline
    ${}^{20}$C & $0^+$ & $18664$ & $1.59$~\cite{Stanoiu:2008zz} & $2.9(3)$ & $1.7^{+\infty}_{-0.5}$
    \\
    ${}^{21}$C & $\frac12^+$ & $-0.014(467)$ & $-0.09^2$ & $-0.07^2$ & $--$ 
    \\
    ${}^{22}$C & $0^+$ & $ 0.11(6)$ & $0.26^2$ & $0.20^2$ &
    \\
    \hline
  \end{tabular}
  \caption{Effective theory parameters, estimates of the expansion parameter, and predicted electric charge radii 
           relative to the core $\delta\left<r^2_\text{E}\right>$
           from Eq.~\eqref{eq:ff_50} for the halo nuclei ${}^{11}$Li, ${}^{14}$Be and ${}^{22}$C. 
           Further explanations are given in the text.}
\label{tab:r2_a}
\end{table}


\begin{figure}[ht]
  \vspace*{0.1cm}
  \centerline{ \includegraphics*[width=0.9\textwidth, angle=0,clip=true]{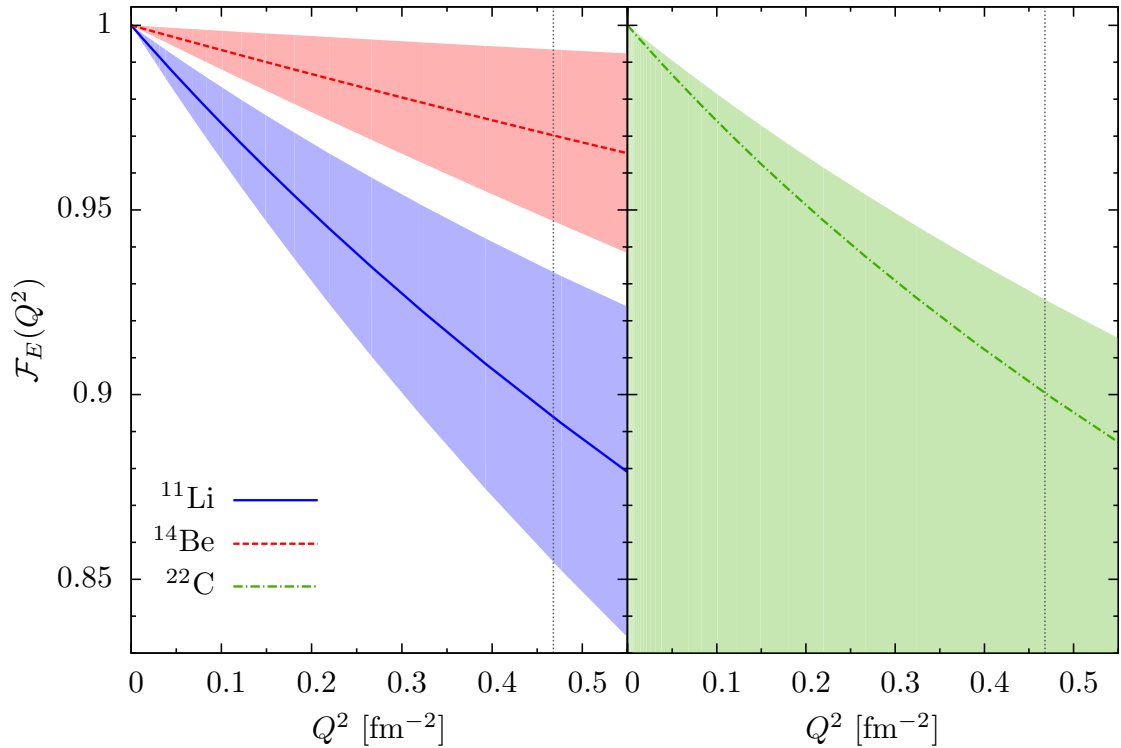} }
  \vspace*{0.0cm}
  \caption{(Color online) The charge form factor $\mathcal F_\text{E}(Q^2)$ for the halo nuclei ${}^{11}$Li (left plot, blue solid line), 
   ${}^{12}$Be (left plot, red long-dashed line) and ${}^{22}$C (right plot, green dash dotted line)
   relative to the core in leading order halo EFT. The estimated theory error for ${}^{11}$Li and  ${}^{12}$Be
   is given by the shaded bands. For ${}^{22}$C, varying the separation energies within their errors gives the shaded region. The vertical dashed lines indicate the breakdown scale from explicit pion exchange.}
  \label{fig:plot_ff_isotopes}
\end{figure}


The ${}^{11}$Li halo nucleus and the ${}^{9}$Li core have both the quantum numbers $J^\pi=\frac32^-$ while
${}^{10}$Li appears to have either $J^\pi=2^-$ or $1^-$. There is some evidence that both S- and P-wave components 
contribute to the neutron halo~\cite{Zhukov-93}. However, we will analyze ${}^{11}$Li under the assumption that
only the S-wave contributes in LO and test the consistency of our assumption with the data. 
P-wave contributions will enter in higher orders. The two-neutron separation energy of 
${}^{11}$Li is $0.37$~MeV and ${}^{10}$Li is $26(13)$~keV above the $n$-$^{9}$Li threshold. 
The first excitation energy of the ${}^{9}$Li ground state is $2.69$~MeV and its one-neutron separation energy is 
$4.06$~MeV. Thus the expansion parameter and the error can be estimated as 
$R_\text{core}/R_\text{halo} \approx \sqrt{B_{cnn}/E^*_c} \approx 0.37$. Calculating the charge radius 
relative to $^9$Li via 
Eq.~\eqref{eq:ff_50} gives $\delta \left<r^2_\text{E}\right> = 1.68(62)$~fm$^2$, where the $\sim40\%$ uncertainty 
comes from the expansion parameter. In Ref.~\cite{Sanchez:2006}, the charge radius was measured with the help of 
high precision laser spectroscopy. The experimental value of 
$\delta\left<r^2_\text{E}\right>_{\rm exp}=1.171(120)$~fm$^2$ is thus compatible with our calculation
within the error bars.


The halo nucleus ${}^{14}$Be and its core ${}^{12}$Be are both in a $J^\pi=0^+$ configuration, while the 
quantum numbers of ${}^{13}$Be are less clear although there is some evidence for $J^\pi=\frac12^-$. For our study, we assume 
that the ${}^{13}$Be dimer has also positive parity. The binding energy of the ${}^{14}$Be trimer is $B_{cnn}=1.27(13)$~MeV 
and the virtual ${}^{13}$Be has $B_{cn}=-510(10)$~keV. The excitation energy of the ${}^{12}$Be core is $E^*_c=2.10$~MeV 
and its one-neutron separation energy is $3.17$~MeV. Thus, the resulting expansion parameter $R_\text{core}/R_\text{halo} 
\approx \sqrt{B_{cnn}/E^*_c} \approx 0.78$ is relatively large. Using Eq.~\eqref{eq:ff_50}, our effective theory  
then predicts a charge radius relative to $^{12}$Be of $\delta\left<r^2_\text{E}\right> = 0.41(32)$~fm$^2$ 
with an $\sim80\%$ error.


There is some theoretical and experimental evidence that ${}^{22}$C is a pure S-wave halo 
nucleus~\cite{Horiuchi:2006,Tanaka:2010}. 
${}^{22}$C and the ${}^{20}$C core both have $J^\pi=0^+$, while ${}^{21}$C is in $J^\pi=\frac12^+$ configuration. 
The two-neutron separation energy $B_{cnn}=0.11(6)$~MeV has a relatively large error. Furthermore, ${}^{21}$C seems 
to be unbound, but $B_{cn}=-0.014(467)$~keV is only poorly known. 
In Ref.~\cite{Stanoiu:2008zz}, a $2^+$ excited state at 1.588(20) MeV above
the ground state was observed.
The one-neutron separation energy of ${}^{20}$C is $2.9(3)$~MeV. We take the central values for $B_{cnn}$ and $B_{cn}$, 
which are also roughly in accord with the allowed parameter region predicted from a recent analysis of the 
matter radius measurement \cite{Tanaka:2010} in the framework of the halo EFT~\cite{Acharya:2013aea}.
Calculating the charge radius relative to $^{20}$C via
Eq.~\eqref{eq:ff_50} gives $\delta\left<r^2_\text{E}\right> = 1.66^{+\infty}_{-0.49}$~fm$^2$, where the uncertainty now comes 
from varying the separation energies within their errors. Due to the poorly known input data, $B_{cn}=B_{cnn}$ is not 
excluded. Since for such values, the charge radius diverges towards positive infinity, the predicted value for 
$\delta\left<r^2_\text{E}\right>$ can only be bound from below, where the lower limit is $1.17$~fm$^2$. For the halo 
nuclei ${}^{14}$Be and ${}^{22}$C, our results are true predictions and can be compared with measurements as soon as the 
corresponding experimental data is available. 


In Fig.~\ref{fig:plot_ff_isotopes}, the charge form factors calculated from Eq.~\eqref{eq:ff_30} are depicted
as a function of the momentum transfer $Q^2$. 
For $Q^2\to\infty$, the form factors vanish. At small momentum transfers, they approach unity as required 
by current conservation. Numerical deviations from unity at vanishing $Q^2$ are less than $10^{-5}$. 
This provides a consistency-check for our calculation. 
For ${}^{11}$Li and  ${}^{12}$Be, the estimated error from higher orders in the effective
theory expansion is given by the shaded bands. For ${}^{22}$C the shaded region originates from varying the binding 
energies within their errors. Our effective theory does neither include explicit pion dynamics nor does it include
the structure of the core. Thus it breaks breaks down for momentum transfers of the order of the pion mass 
$m_\pi^2\approx0.5$~fm$^{-2}$ 
as indicated by the vertical dashed lines in  Fig.~\ref{fig:plot_ff_isotopes}.

                                                                                     
\begin{figure}[ht]
  \centerline{ \includegraphics*[width=0.8\textwidth, angle=0,clip=true]{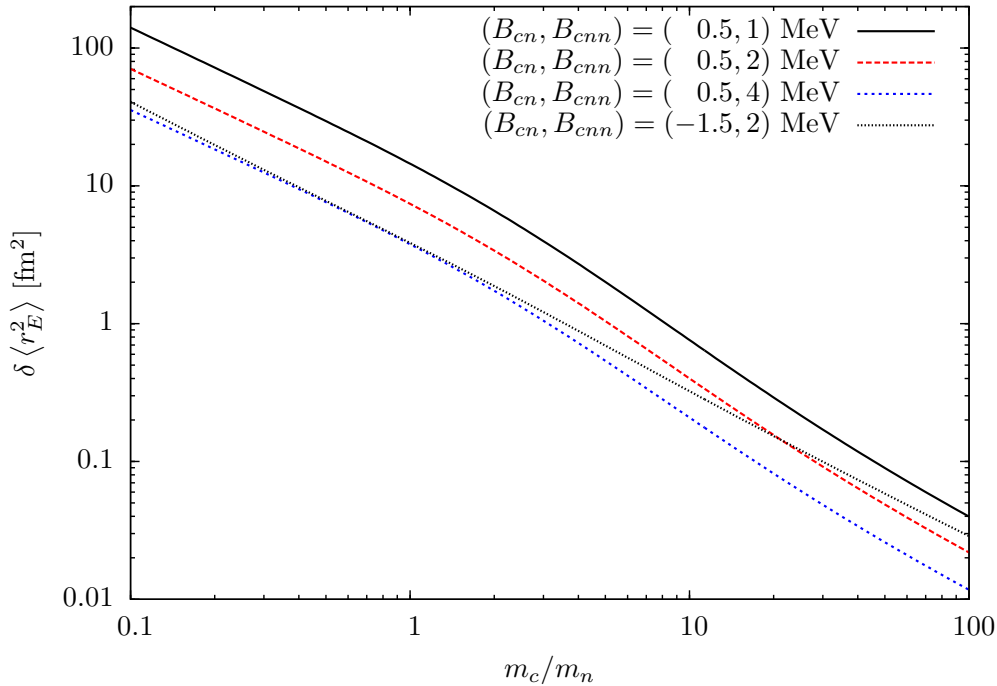} }
  \vspace*{0.0cm}
  \caption{(Color online) The electric charge radius relative to the core
   $\delta\left<r^2_\text{E}\right>$ as a function of mass ratio $m_c/m_n$ 
   for different binding energies $B_{cn}$ and $B_{cnn}$.}
  \label{fig:cnn_r2__mass_ratio}
\end{figure}
\begin{figure}[ht]
  \centerline{ \includegraphics*[width=0.85\textwidth, angle=0,clip=true]{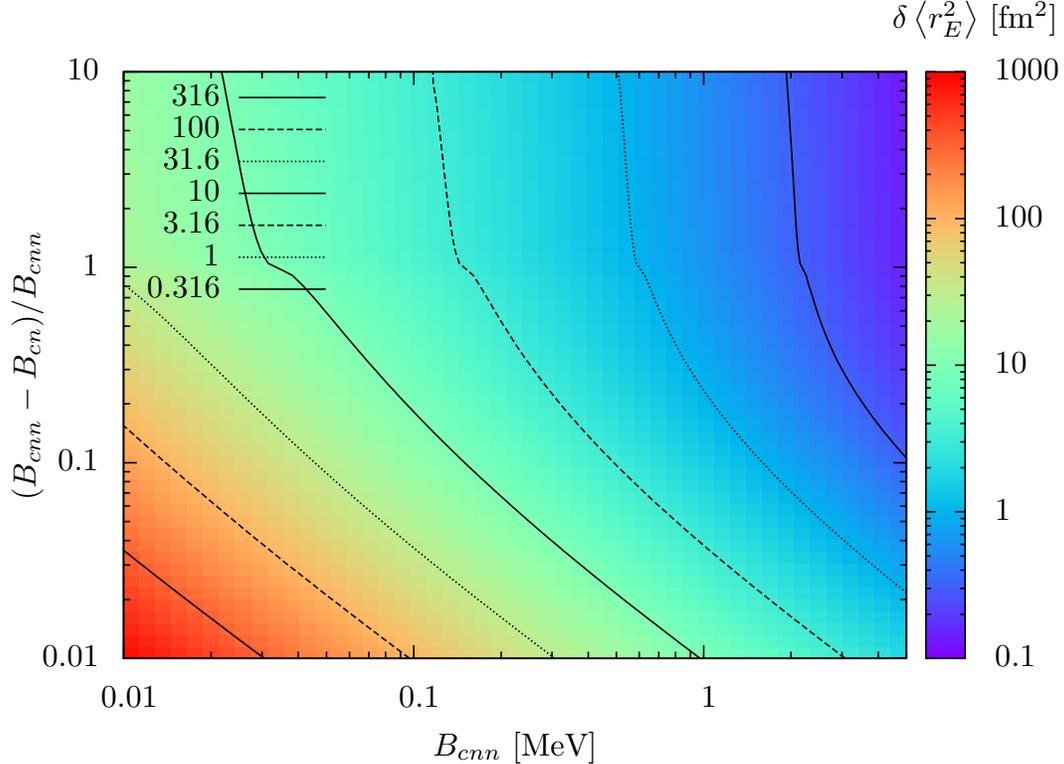} }
  \vspace*{0.0cm}
  \caption{(Color online) The electric charge radius relative to the core
 $\delta\left<r^2_\text{E}\right>$ as a function of the energies 
 $B_{cnn}$ and $1-B_{cn}/B_{cnn}$ for fixed mass ratio $m_c/m_n = 10$.}
  \label{fig:cnn_r2__energies_contour}
\end{figure}


Another interesting aspect is to understand the general dependencies of the charge radius on the core mass 
and the $cn$ and $cnn$ binding energies. In Fig.~\ref{fig:cnn_r2__mass_ratio}, we therefore plot 
$\delta\left<r^2_\text{E}\right>$ as a function of the mass ratio $m_c/m_n$ for fixed energy values $B_{cnn}$ and $B_{cn}$. 
The charge radius is always positive and decreases for growing core mass $m_c$. 
This reflects that $\delta\left<r^2_\text{E}\right>$ for  the two-neutron halo system 
originates from the recoil effect of the charged core. For core masses 
below $2m_n$ the radius roughly falls of like $\sim1/m_c$. Around $m_c\approx 2m_n$, the slope changes and
the charge radii drop even faster as $m_c/m_n$ increases.


In Fig.~\ref{fig:cnn_r2__energies_contour}, the dependence of the charge radius relative to the core
on the energies $B_{cnn}$ and $1-B_{cn}/B_{cnn}$
is shown for a fixed mass ratio $m_c/m_n = 10$. 
The region $1-B_{cn}/B_{cnn}<1$ corresponds to a bound $cn$-system, while
$1-B_{cn}/B_{cnn}>1$ implies that the $cn$-system is unbound. 
In this region the $cnn$-system is Borromean. If $1-B_{cn}/B_{cnn}
\ll 1$ the $cn$ system is deeply bound and the three-body problem 
reduces to a two-body problem of $cn$ and another neutron.
As one would also naively expect, $\delta\left<r^2_\text{E}\right>$ 
grows as both the 
binding energies for the three-body system $B_{cnn}$ and the binding energy of the dimer-particle system 
$B_{cnn}-B_{cn}$ decrease. However, the exact functional dependencies on the 3 quantities $m_c$, $B_{cnn}$ and 
$B_{cn}$  are more complicated. Also note that there is a sudden increase in $\delta\left<r^2_\text{E}\right>$  
along the line $1-B_{cn}/B_{cnn}=1$ where
the $cnn$-system becomes Borromean. This leads to a ridge 
along $(B_{cnn}-B_{cn})/B_{cnn}=1$ that is most easily seen in the contour lines.

\section{Summary and Conclusion}
\label{sec:conclusion}


In this paper, we have set up a formalism to describe the electromagnetic structure of $2n$ halo nuclei 
within halo EFT. We assumed that the $cn$ and $nn$ S-wave scattering lengths are much larger than the range
of the interaction and calculated the form factors and charge radii of various halo nuclei to leading 
order in the expansion in $R_\text{core}/R_\text{halo}$. The renormalization of the trimer propagator
and the extraction of the trimer wave function renormalization were discussed in detail.
The charge form factor receives contributions from three
different classes of diagrams illustrated in Fig.~\ref{fig:FG_ff_contributions}.
In all three contributions, the irreducible trimer-dimer-particle three-point 
function $\vec {\mathcal G}^\text{irr}$ from Eq.~\eqref{eq:ftc_40} appears naturally. 
Current conservation insures the correct normalization of the charge form
factor $\mathcal F_\text{E}(0)=1$. Numerically, we find deviations from unity at vanishing $Q^2$ 
of less than $10^{-5}$ in our calculation. 


We have applied our formalism to ${}^{11}$Li, ${}^{14}$Be and ${}^{22}$C 
calculated their charge form factors and radii relative to the core to leading order in
$R_\text{core}/R_\text{halo}$.  The resulting charge radii are 
$\delta\left<r^2_\text{E}\right>_{{}^{11}\text{Li}} = 1.68(62)$~fm$^2$, 
$\delta\left<r^2_\text{E}\right>_{{}^{14}\text{Be}} = 0.41(32)$~fm$^2$, and 
$\delta\left<r^2_\text{E}\right>_{{}^{22}\text{C}} = 1.66^{+\infty}_{-0.49}$~fm$^2$. 
For ${}^{11}$Li a comparison with the measured value $1.171(120)$~fm$^2$ shows good agreement within the $\sim40\%$ 
uncertainty originating from the expansion parameter of our leading order calculation. 
The other charge radii are true predictions that can be compared to future experiments.
For a more quantitative comparison with experiment, the extension to higher orders is clearly required. 
This includes the treatment of effective range effects~\cite{Kaplan:1996nv,Hammer:2001gh,Ji:2011qg}
as well as of P-wave interactions in ${}^{11}$Li and ${}^{14}$Be. Electromagnetic breakup reactions
can reveal additional information on the structure of $2n$ halo nuclei. An investigation
of this process in the framework of halo EFT is in progress~\cite{AHHP:2013}.


Finally, we have investigated the dependence of the charge radius on the core mass and the 
$cnn$ and $cn$ binding energies. Our results are summarized in Figs.~\ref{fig:cnn_r2__mass_ratio} and 
\ref{fig:cnn_r2__energies_contour}. As expected, the charge radius decreases with increasing core radius.
However, the exact dependence for large core masses deviates from a simple $1/m_c$ dependence.
Moreover, the charge radius increases as binding energies $B_{cnn}-B_{cn}$ and $B_{cnn}$ decrease. In particular,
we found  a sudden increase of the charge radius along the line $B_{cn}=0$ where
the $cn$-system becomes Borromean. A better understanding of these
characteristics will require further studies.

We note that our approach predicts not only the radii but also the full
charge form factor of the halos. To date, electron scattering experiments 
which would give access to the charge form factor have not been carried out. 
Such experiments are planned at FAIR (ELISe) \cite{ELIse}. 
However, ELISe is not part of the start version of FAIR and corresponding 
experiments are far in the future.


\begin{acknowledgments}
We thank D.R.~Phillips for discussions and B.~Acharya for comments on
the manuscript. This work was supported in part by
the DFG and the NSFC through funds provided to the Sino-German CRC 110 
``Symmetries and the emergence of structure in QCD'', by the BMBF under 
contract 05P12PDFTE, and by the US Department of Energy, Office of Nuclear
Physics, under Contract No. DE-AC02-06CH11357.

\end{acknowledgments}

\appendix

\section{Form factor contributions}
\label{sec:appedix}

\noindent


In this appendix, we explicitly derive the three different contributions to the charge form factor
discussed in Sec.~\ref{sec:electric_formfactors}. Therefore, we first give an expression for $i\Gamma_0(\textbf{Q})$ from 
Eq.~\eqref{eq:ff_15} as the sum of all irreducible Feynman diagrams with external trimer lines and a photon coupled to the core. 
It can be conveniently written as
\begin{equation}
  \begin{split}
    i\Gamma_0(\textbf{Q}) \ =& \ \underset{|\textbf{p}|<\Lambda}{\int}\frac{\text{d}^4p}{(2\pi)^4}  
\underset{|\textbf{k}|<\Lambda}{\int}\frac{\text{d}^4k}{(2\pi)^4}
    \\
    & \quad \times \ i{\vec {\mathcal G}^\text{irr}}(E,\textbf{P}, p_0,\textbf{p})^\text{T} \ i \bar \Gamma_0(E,\textbf{P}, p_0, 
\textbf{p}, \textbf{K},k_0,\textbf{k}) \  i {\vec {\mathcal G}^\text{irr}}(E,\textbf{K}, k_0, \textbf{k}) \ .
  \end{split}
  \label{eq:app_03}
\end{equation}
The quantity ${\mathcal G}^\text{irr}_i(E,\textbf{P}, p_0,\textbf{p})$ is the irreducible trimer-dimer-particle three-point function in 
general kinematics, where the trimer, dimer and particle four-momenta are written as $P^\mu$, $\frac{M_i}{M_\text{tot}}P^\mu + p^\mu $ 
and $\frac{m_i}{M_\text{tot}} P^\mu - p^\mu $, respectively. The energy reads $E=P_0-\textbf{P}^2/(2M_\text{tot})$, where the kinetic 
energy of the three-body bound state is subtracted. ${\mathcal G}^\text{irr}_i(E,\textbf{P}, p_0,\textbf{p})$ is related to the 
center-of-mass quantity ${\mathcal G}^\text{irr}_i(E,p)$ from Eq.~\eqref{eq:ftc_40} through the integral equation
\begin{equation}
  \begin{split}
    \mathcal G^\text{irr}_{i}(E,\textbf{P}, p_0,\textbf{p}) \ =& \ \sum_{j=0}^1 \int_0^\Lambda \text{d}q \ R_{ij}
\Bigl(\frac{M_i}{M_\text{tot}}E + p_0 - \frac{ \textbf{P} \cdot \textbf{p} }{ M_\text{tot}} + \frac{ \textbf{p}^2 }{ 2m_i }, 
p, q\Bigr) \Big|_{H=0}
    \\
    & \qquad \qquad \times \ \bar D_j(E, q) \, \mathcal G^\text{irr}_j(E,q) \ + \ \delta_{0i} \ .
  \end{split}
  \label{eq:app_06}
\end{equation}
The matrix valued function $i\bar \Gamma_0$ in Eq.~\eqref{eq:app_03} is the sum of the three diagrams that are depicted in the 
lower row in Fig.~\ref{fig:FG_ff_contributions}. Its contributions (a), (b), and (c) are products of the corresponding 
delta-functions, propagators and couplings.


Since in the Breit frame in- and outgoing three-body bound states have $E=-B$, we drop this redundant energy variable 
in ${\vec {\mathcal G}^\text{irr}}$ and $\bar{D}$ in all subsequent equations.


\subsection{Contribution $\mathcal F^{(a)}$}

\noindent


\begin{figure}[ht]
  \centerline{ \includegraphics*[angle=0,clip=true]{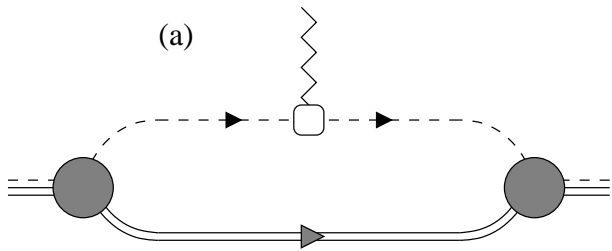} }
  \vspace*{0.0cm}
  \caption{Type-(a) contribution to the form factor matrix element where the photon couples to a core field 
   propagating parallel to the $d_0$-dimer.
    Notation is as in Fig.~\ref{fig:FG_2ftc_full_G}.}
  \label{fig:FG_app_contribution_a}
\end{figure}


We start with the type-(a) contribution depicted in Fig.~\ref{fig:FG_app_contribution_a}
where the photon couples to a core field propagating parallel to a $d_0$-dimer. 
All non-perturbative physics is
contained in the irreducible trimer-dimer-particle three-point function ${\vec {\mathcal G}^\text{irr}}$.
One of the two four-momentum integrations in Eq.~\eqref{eq:app_03} is absorbed by a delta-function and the remaining 
loop four-momentum $q^\mu$ can be chosen in such a way that the resulting expression is symmetric under $\textbf{q} \mapsto -\textbf q$.
The $q^0$ integration can then be performed analytically, leading to two contributions from picking the poles 
of the two core propagators. The trimer-dimer-particle three-point functions ${\mathcal G}^\text{irr}$
appear in off-shell kinematics and are related to on-shell center-of-mass quantities via Eq.~\eqref{eq:app_06} at the expense of 
two additional momentum integrations.
Choosing spherical coordinates for $\textbf q$, with $\textbf Q=\textbf K - \textbf P$ pointing in 
the $z$-direction, the azimuthal dependence can also be integrated out. We are then left with a remaining integral over $q=|\textbf q|$ 
and the polar angle $x=\cos(\angle(\textbf Q, \textbf q))$.  Since the $d_0$ dimer is propagating 
parallel to the core, the integral equation \eqref{eq:ftc_45} leads to extra contributions from the three-body force.
Finally, we end up with
\begin{equation}
  \begin{split}
    \mathcal F^{(a)}( Q^2 ) \ =& \  Z_\text{tr}\,H^2 \Biggl\{ \int_0^\Lambda \text{d}p \  \int_0^\Lambda \text{d}k \ 
{\vec {\mathcal G}}^\text{irr}(p)^\text{T} \, \bar{D}(p) \ \Upsilon^{(a)}(Q,p,k)  \ \bar{D}(k) \, {\vec {\mathcal G}}^\text{irr}(p)
    \\
    & \qquad \qquad + \ 2\int_0^\Lambda \text{d}p \  {\vec {\mathcal G}}^\text{irr} (p)^\text{T} \, \bar{D}(p) \ 
{ {\vec \Upsilon}^{(a)}(Q,p) }
    \ + \ \Upsilon^{(a)}_0(Q) \Biggr\}\ , 
  \end{split}
  \label{eq:app_09}
\end{equation}
where the matrix-, vector- and scalar-valued functions $\Upsilon^{(a)}(Q,p,k)$, ${\vec \Upsilon}^{(a)}(Q,p)$ and $\Upsilon^{(a)}_0(Q)$, 
using the shortened notation $\Upsilon^{(a)}_{\dots}(Q\dots)$, are given by:
\begin{equation}
  \begin{split}
    \Upsilon^{(a)}_{\dots}(Q\dots) \ =& \ \frac{\tilde \mu_0}{4} \int_0^\Lambda \frac{\text{d}q}{q} \, \int_{-1}^1 \frac{\text{d}x}{x} \ 
\chi^{(a)}_{\dots} \left( \frac{M_0}{M_\text{tot}} \frac{Q}{2},q,x \dots \right)\ ,
\end{split}
\end{equation}
where
\begin{equation}
\begin{split}
    \chi^{(a)}_{ij}(r,q,x,p,k) \ =& \ \frac{1}{r} \Biggl\{ R_{i0}(-B + 2\frac{q r x}{\tilde \mu_0}, p , d(r,q,x) )\Big|_{H=0} \ 
\bar{D}_0 \left( -B -\frac{ m_0 \, r^2 }{2\tilde \mu_0^2} + \frac{ q r x}{\tilde \mu_0}, q \right)
    \\ 
    & \qquad \times \ R_{0j}(-B, d(r,q,-x), k )\Big|_{H=0}
    \\
    & - \ R_{i0}(-B , p , d(r,q,x) )\Big|_{H=0} \ \bar{D}_0 \left( -B - \frac{m_0 \, r^2}{2\tilde \mu_0^2} - \frac{ q r x}{\tilde \mu_0}, 
q \right)
    \\
    & \qquad \times \ R_{0j}(-B - 2\frac{q r x}{\tilde \mu_0}, d(r,q,-x),k )\Big|_{H=0} \Biggr\}\ ,
    \\
    \chi^{(a)}_i(r,q,x,p) \ =& \  \frac{1}{r} \Biggl\{ R_{i0}(-B + 2\frac{q r x}{\tilde \mu_0}, p , d(r,q,x))\Big|_{H=0} \ 
 \bar{D}_0 \left( -B -  \frac{ m_0 \, r^2}{2\tilde \mu_0^2} + \frac{ q r x}{\tilde \mu_0}, q \right)
    \\ 
    & - R_{i0}(-B , p , d(r,q,x) )\Big|_{H=0} \ \bar{D}_0 \left( -B - \frac{m_0 \, r^2}{2\tilde \mu_0^2} - \frac{ q r x}{\tilde \mu_0}, 
q \right) \Biggr\}\ ,
    \\
    \chi^{(a)}_0(r,q,x) \ =& \  \frac{1}{r} \Biggl\{ \bar{D}_0 \left( -B - \frac{m_0\, r^2}{2\tilde \mu_0^2} + \frac{ q r x}{\tilde \mu_0}, 
q \right) \ - \  \bar{D}_0 \left( -B - \frac{m_0 \, r^2}{2\tilde \mu_0^2} - \frac{ q r x}{\tilde \mu_0}, q \right) \Biggr\}\ ,
 \end{split}
  \label{eq:app_10}
\end{equation}
with $d(r,q,x) = \sqrt{q^2 + 2 qrx + r^2}$.
One can easily check that the symmetry $\Upsilon^{(a)}(Q, p,k) = \Upsilon^{(a)}(Q, k,p)^\text{T}$ holds as required. 
Note also that the limit $Q\to0$ in Eq.~\eqref{eq:app_10} exists, but prefactors $\propto 1/Q$ cause numerical 
instabilities for very small momentum transfer.

\subsection{Contribution $\mathcal F^{(b)}$}

\noindent


\begin{figure}[ht]
  \centerline{ \includegraphics*[angle=0,clip=true]{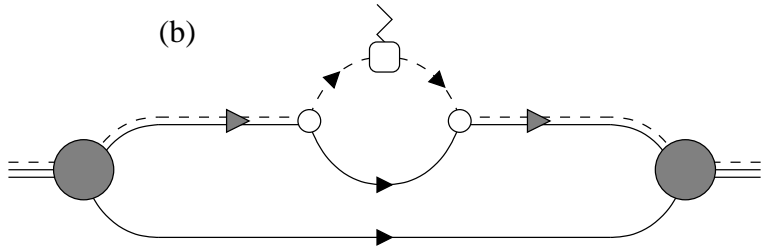} }
  \vspace*{0.0cm}
  \caption{Type-(b) contribution to the form factor matrix element where the photon couples to a
    core field inside a $nc$ bubble. Notation is as in Fig.~\ref{fig:FG_2ftc_full_G}.}
  \label{fig:FG_app_contribution_b}
\end{figure}


We now consider  contributions to the formfactor of type (b),
where the photon couples to a core field inside a $nc$ bubble.
The corresponding Feynman diagram is depicted in Fig.~\ref{fig:FG_app_contribution_b}.
The bubble sub-diagram can be calculated analytically using Feynman integrals. Analogue to case~(a) one of the two four-momentum 
integrations in Eq.~\eqref{eq:app_03} is absorbed by a delta-function and the remaining loop 
four-momentum $q^\mu$ can be chosen in a symmetric way. Applying Eq.~\eqref{eq:app_06} then again leads to two additional momentum 
integrations. In spherical coordinates, the azimuthal 
integration can then be performed leading to two remaining integrals over $q=|\textbf q|$ and the polar 
angle $x=\cos(\angle(\textbf Q, \textbf q))$. Since diagrams of type (b) only appear in channel 1 where the neutron-core dimer 
is present, there are no three-body force contributions in the integral. We find
\begin{equation}
  \mathcal F^{(b)}(Q^2) \ = \ Z_\text{tr}\,H^2 \ \int_0^\Lambda \text{d}p \int_0^\Lambda \text{d}k \ 
{\vec {\mathcal G}}^\text{irr}(p)^\text{T} \  \bar{D}(p) \ \Upsilon^{(b)}(Q , p,k) \ \bar{D}(k) \ \vec {\mathcal G}^\text{irr}(k) \ ,
  \label{eq:app_19}
\end{equation}
where the matrix-valued function $\Upsilon^{(b)}(Q , p,k)$ is given through:
\begin{equation}
  \begin{split}
    \Upsilon^{(b)}_{ij}(Q, p,k) \ =& \ \ \frac{\mu_1 \, |a_1| }{(2\pi)^2} \, \frac{m_0m_1}{2M_\text{tot} } \ \int_{0}^\Lambda \text{d}q \, 
q^2 \, \int_{-1}^1 \text{d}x \ \chi^{(b)}_{ij} \left( \frac{m_1}{M_\text{tot}} \frac{Q}{2}, q,x, p,k\right)\ ,
    \\
    \chi^{(b)}_{ij}(s,q,x, p,k) \ =& \ \frac{ R_{i1}(-B,p,d(s,q,-x))\Big|_{H=0} }{ \frac{1}{a_1} - \sqrt{ b_1(-B,d(s,q,-x) ) } } \ 
\frac{1}{s} \ \Biggl\{ \arctan\left( \frac{   \frac{M_\text{tot}}{M_1} s + \frac{m_0}{M_1} q x }{ \sqrt{ b_1(-B,d(s,q,-x) ) }  } \right)
    \\
    + \ &\arctan \left( \frac{ \frac{M_\text{tot}}{M_1} s - \frac{m_0}{M_1} q x }{ \sqrt{ b_1(-B,d(s,q,x) ) }  } \right) \Biggr\} \ 
\frac{ R_{1j}(-B,d(s,q,x),k)\Big|_{H=0} }{ -\frac{1}{a_1} + \sqrt{ b_1(-B,d(s,q,x) ) } } \quad . 
  \end{split}
  \label{eq:app_20}
\end{equation}
Again $\Upsilon^{(b)}(Q, p,k) = \Upsilon^{(b)}(Q, k,p)^\text{T}$ holds
and the limit $Q\to0$ can lead to numerical instabilities.

\subsection{Contribution $\mathcal F^{(c)}$}

\noindent


\begin{figure}[ht]
  \centerline{ \includegraphics*[angle=0,clip=true]{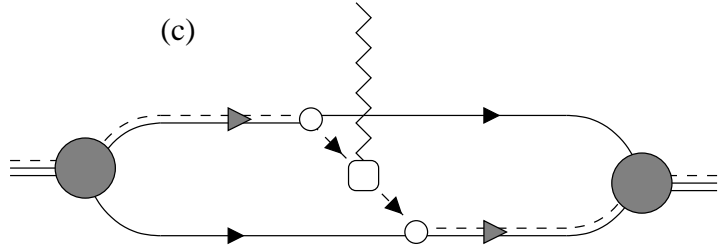} }
  \vspace*{0.0cm}
  \caption{Type-(c) contribution to the form factor matrix element where the photon couples to
          a core field exchanged between $\vec{d}_1$ dimers. Notation is as in Fig.~\ref{fig:FG_2ftc_full_G}.}
  \label{fig:FG_app_contribution_c}
\end{figure}


In the remaining contribution of type (c) shown in Fig.~\ref{fig:FG_app_contribution_c},
the photon couples to a core field that is exchanged  between $\vec{d}_1$ dimers.
For both loops, the energy integrals can be performed analytically, leading to on-shell conditions for the
trimer-dimer-particle three-point function ${\mathcal G}^\text{irr}$. In spherical coordinates, one of the two azimuthal 
integrals can be solved analytically, such that in the end five integrals remain. For $\mathcal F^{(c)}$, we then get
\begin{equation}
  \mathcal F^{(c)}(Q^2) \ = \  Z_\text{tr}\,H^2 \ \int_0^\Lambda \text{d}p \int_0^\Lambda \text{d}k \ 
{\vec {\mathcal G}}^\text{irr}(p)^\text{T} \ \bar{D}(p) \ \Upsilon^{(c)}(Q , p,k) \ \bar{D}(k) \ \vec {\mathcal G}^\text{irr}(k)\ ,
  \label{eq:app_29}
\end{equation}
with a the matrix-valued function $\Upsilon^{(c)}(Q , p,k)$ given by:
\begin{equation}
  \begin{split}
    \Upsilon^{(c)}_{ij}(Q, p,k) \ =& \ \frac{2}{|a_1|} \ \int_{-1}^1 \text{d}x \int_{-1}^1 \text{d}y  \int_0^\pi \text{d}\phi \ 
\chi^{(c)}_{ij} \left( \frac{m_1}{M_\text{tot}} Q, x,y,\phi, p,k\right)\ ,
    \\
    \chi^{(c)}_{ij}(t, x,y,\phi, p,k) \ =& \ \delta_{i1} \, \delta_{1j} 
    \\
    \times \ \Biggl[ p^2 + k^2 + t^2 - &2\left( ky + \frac{m_1}{M_1} p x \right) t + 2 \frac{m_1}{M_1} pk \left[ 
\sqrt{1-x^2}\sqrt{1-y^2} \cos\phi + xy \right] +  2\mu_1B \Biggr]^{-1}
    \\
    \times \ \Biggl[ p^2 + k^2 + t^2 + &2\left( px + \frac{m_1}{M_1} k y \right) t + 2 p \frac{m_1}{M_1}k \left[ 
\sqrt{1-x^2}\sqrt{1-y^2} \cos\phi + xy \right] +  2\mu_1B \Biggr]^{-1}
  \end{split}
  \label{eq:app_30}
\end{equation}
As in the other cases, the symmetry condition $\Upsilon^{(c)}(Q, p,k) = \Upsilon^{(c)}(Q, k,p)^\text{T}$ holds.


The dependence of all three formfactor contributions (a), (b), and (c) on $Q^2$ only
is not directly evident from Eqs.~\eqref{eq:app_09},~\eqref{eq:app_19} and~\eqref{eq:app_29}
but numerically it is satisfied. Moreover, the charge form factor  $\mathcal F(Q^2)$ is automatically
normalized to unity at zero momentum transfer.

\end{document}